\documentclass[%
amsmath,amssymb,
 aps,
prl,twocolumn,showpacs
]{revtex4-1}
\usepackage{graphicx}
\usepackage{bm}
\usepackage{hyperref}
\usepackage{color}
\usepackage[normalem]{ulem}
\usepackage{units}

\begin{document}

\title{Observation of Multimode Strong Coupling of Cold Atoms to a 30-m Long Optical Resonator}

\author{Aisling Johnson$^1$, Martin Blaha$^1$, Alexander E. Ulanov$^2$, Arno Rauschenbeutel$^{1,3}$, Philipp Schneeweiss$^1$, and J\"urgen Volz$^{1}$}
\email{juergen.volz@tuwien.ac.at}
\affiliation{$^1$Vienna Center for Quantum Science and Technology, TU Wien-Atominstitut, Stadionallee 2, 1020 Vienna, Austria,}%

\affiliation{$^2$Russian Quantum Center, 100 Novaya St., Skolkovo, Moscow 143025, Russia,}

\affiliation{$^3$ Department of Physics, Humboldt-Universit\"at zu Berlin, Berlin, Germany.}

\date{\today}

\begin{abstract}{We report on the observation of multimode strong coupling of a small ensemble of
atoms interacting with the field of a 30--m long fiber resonator containing a nanofiber section. The collective light--matter coupling strength exceeds the free spectral range and the atoms couple to consecutive longitudinal resonator modes. The measured transmission spectra of the coupled atom--resonator system provide evidence of this regime, realized with a few hundred atoms with an intrinsic single-atom cooperativity of 0.26. These results are the starting point for studies in a new setting of light--matter interaction, with strong quantum non--linearities and a new type of dynamics.}
\end{abstract}

\maketitle

In recent years, the control of light-matter interaction at the fundamental quantum-mechanical level has already proven successful for, e.g, the implementation of quantum metrology, quantum simulation, quantum communication, and information processing. A milestone in this endeavor was the realization of strong coupling between single quantum emitters and single photons in CQED \cite{Kimble1998}; in this regime, the coherent emitter-photon dynamics is faster than all incoherent decay rates of the system. In ensuing experiments, this achievement led to a variety of applications~\cite{Reiserer2015} such as controlled quantum state transfer~\cite{Wilk2007, Boozer2007, Stute2013}, back-action-free state detection~\cite{Volz2011} or the implementation of new quantum protocols~\cite{Tiecke2014,Hacker2016,Scheucher2016}.

Interfacing the resonator mode simultaneously with many identical quantum emitters such as neutral atoms not only allowed to reach extremely large collective light-matter coupling strengths~\cite{Colombe2007}, but also initiated new approaches for quantum devices. These include single photon transistors~\cite{Chen2013}, new methods for the generation of collective mesoscopic quantum states~\cite{Leroux2010, Haas2014} and real-time observation of optical photons~\cite{Hosseini2016}. Furthermore, in such systems, the resonator mode can be employed to provide an infinite-range interaction between the different atomic qubits, a key ingredient for the study of the quantum dynamics of collective light-matter systems~\cite{Baumann2010,Ritsch2013,Kessler2014,Spethmann2016}. Moving beyond coupling with a single electromagnetic mode, new experimental platforms where atoms couple with a number of degenerate, higher-order transverse resonator modes, are under development~\cite{Schine2016,Ballantine2017,Jia2018}. These could be employed, for instance, for photonic quantum information processing~\cite{Wickenbrock2013}.

A largely unexplored regime of cavity QED is reached when a single emitter or an ensemble of emitters interacts strongly with several non-degenerate, longitudinal modes of a long cavity. This scenario bridges a gap between two archetypical regimes of light--matter interaction: the strong coupling of emitters to a single resonantly enhanced field mode, giving rise to, e.g., vacuum-induced Rabi-oscillations~\cite{Brune1996}, and the coupling of emitters to a single spatial mode with a continuous spectrum, enabling applications such as efficient single photon sources~\cite{Lodahl2015}. In this regime of so-called multimode strong coupling, the emitter--resonator coupling strength exceeds the free-spectral range (FSR) of the resonator as well as the single-emitter decay rate. Previous investigations in the microwave and acoustic domains \cite{Sundaresan2015, Moores2018} have recently put this novel regime, also coined superstrong-coupling regime \cite{Meiser2006}, in the spotlight. 

Here, we realize multimode strong coupling (MMSC) with a small ensemble of atoms that is coupled to a 30-m long optical fiber ring resonator.  In the transmission spectrum, several longitudinal cavity resonances are significantly modified and coupling strengths of up to twice the FSR are observed. We determine the number of coupled atoms by analysing the second order correlation function of fluorescence light scattered into the fiber ring. Remarkably, the MMSC regime is reached in our experiment with as little as 200 atoms. At the same time, we infer an intrinsic single-atom cooperativity of 0.26, meaning that the system's dynamics depends on a quantized degree of freedom, as its response will be nonlinear at the level of a few photons. This sets our work apart from more conventional situations in, e.g., laser physics, where the dispersion of a macroscopic medium inside the resonator can exceed the FSR.

Our experiment relies on four key aspects~\cite{Schneeweiss2017}. First, the light field in the resonator and the atoms are coupled via the evanescent field surrounding the nanofiber-waist of a tapered optical fiber (TOF)~\cite{Solano2017, Nieddu2016}. The strong transverse confinement of the nanofiber-guided light gives rise to a large single-pass atom--field coupling strength. Second, thanks to the low transmission losses, a TOF can be turned into an optical resonator. Indeed, strong atom--light coupling was  demonstrated for TOF-based Fabry-Perot and ring resonators \cite{Kato2015, Ruddell2017}. Third, our resonator consists of a $\sim 30$~m long fiber ring, whose FSR is as small as $\nu_{\textrm FSR}=7.1$~MHz, eight orders of magnitude smaller than the carrier frequency of the optical field. In this geometry, the single-atom cooperativity $C_1 = g_1^2/2(\kappa_0 + \kappa_{\text{ext}})\gamma$ ($\kappa_0$ being the intrinsic cavity loss rate, $\kappa_{\text{ext}}$ the in- and out-coupling rate, $\gamma$ the spontaneous emission rate of the atom, and $g_1$ the single-atom, single-photon coupling strength) is  independent of resonator length as long as the fiber propagation losses can be neglected~\cite{Schneeweiss2017}. Under this condition, the FSR of the resonator can be decreased without affecting $C_1$. Finally, to further enhance the coupling, we use a small ensemble of $N$ laser-cooled atoms, interacting with the resonator with a collective strength $g_N = \sqrt{N} g_1$.

\begin{figure}[h!]
\centerline{\includegraphics[width=1\columnwidth]{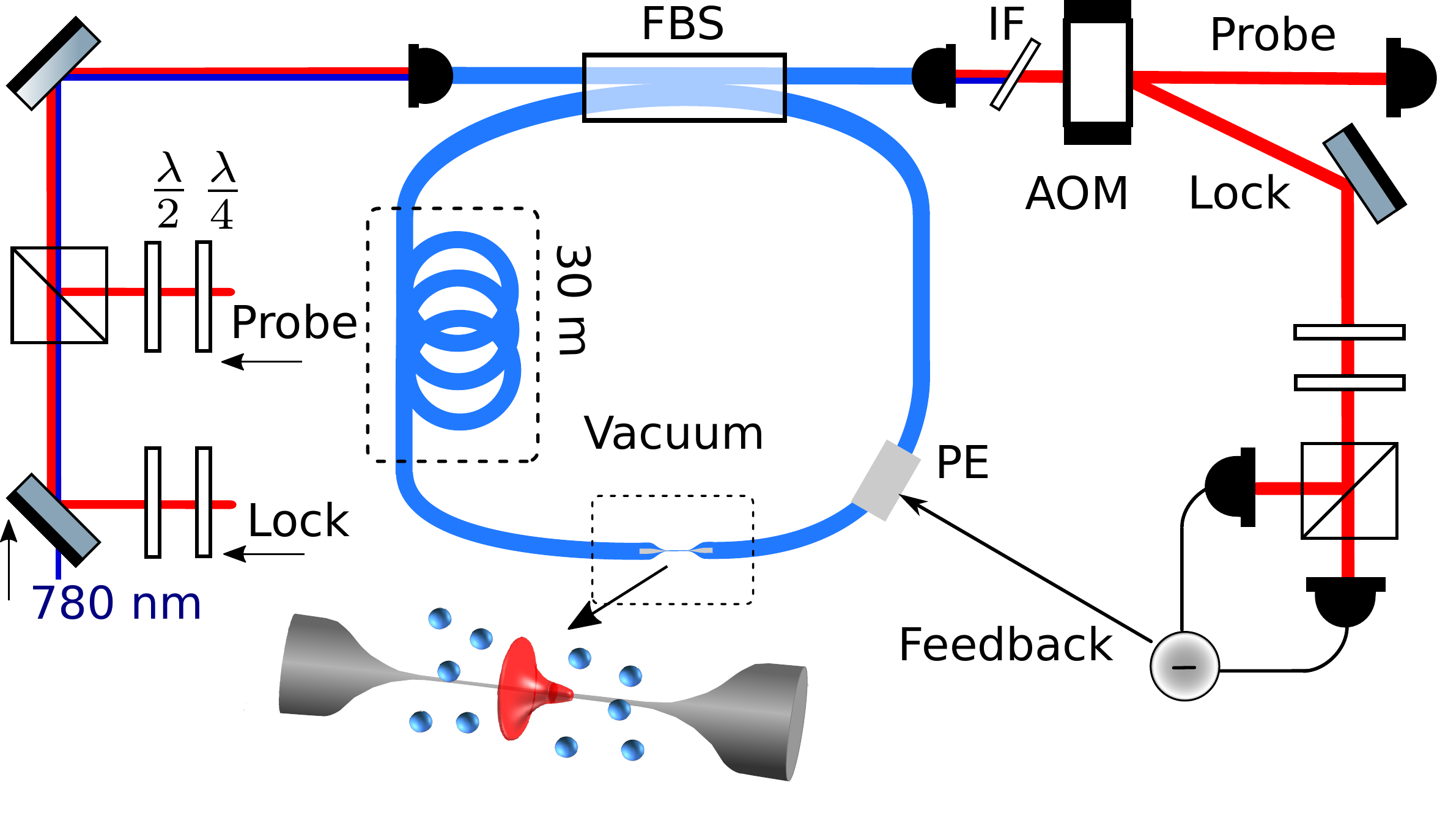}}
\caption{Sketch of the 30-m long fiber-ring resonator with a nanofiber section for coupling atoms to the resonator mode via its evanescent field part. The probe light (optical power: 1~pW, wavelength: 852~nm, resonant with the D2 cycling transistion of cesium), the lock light (10~$\mu$W), and the heating laser light (1~mW, wavelength: 780~nm) are combined, fiber coupled, and sent into a variable fiber beamsplitter (FBS) which couples light into and out of the resonator. The polarization of the probe and lock light are independently controlled. The resonator is placed inside an acoustically and thermally insulating box, except for the section fed into the vacuum chamber, which contains the nanofiber-waist of the TOF (diameter: 400~nm, length: 1~cm). A cloud of laser-cooled atoms is overlapped with the waist. During the preparation phase, the lock light is directed towards a polarization detection set-up using an acousto-optical modulator (AOM). The generated lock signal is fed back onto a piezoelectric element (PE) which stretches the fiber to compensate for external perturbations. During the probing phase, the resonator output is then redirected onto a single-photon counting module. An interference filter (IF) removes the 780-nm heating light prior to the detection.
}
\label{fig.setup}
\end{figure}

The experimental sequence alternates between two phases: the preparation phase, during which the resonator frequency is actively stabilized and a cloud of cold Cs is prepared around the nanofiber by means of a magneto-optical trap (MOT), and a probing phase, where the locking is intermittently interrupted. The resonator frequency is locked to the probe laser using polarization spectroscopy with feedback onto a piezoelectric actuator (see Supplemental Material, \cite{SuppMat}) \cite{Hansch1980, Libson2015}. In our setup, see Fig.~\ref{fig.setup}, a variable fiber beamsplitter (FBS) allows us to control the coupling rate of light into and out of the resonator~\cite{Schneeweiss2017}. We use two settings of this FBS: For the first setting, the coupling rate is chosen such that the on-resonance transmission through the coupling fiber is minimized for the bare resonator. We refer to this critical coupling configuration as the fiber-loop being closed. All later measurements on MMSC are performed under this condition. For the second setting, the light only takes one round-trip in the resonator. We also refer to this fully overcoupled regime as the fiber-loop being open. This second setting allows us to deduce the single-pass on-resonance optical depth (OD) of our atomic sample measured through the waveguide.

\begin{figure}[h!]
\centerline{\includegraphics[width=1\columnwidth]{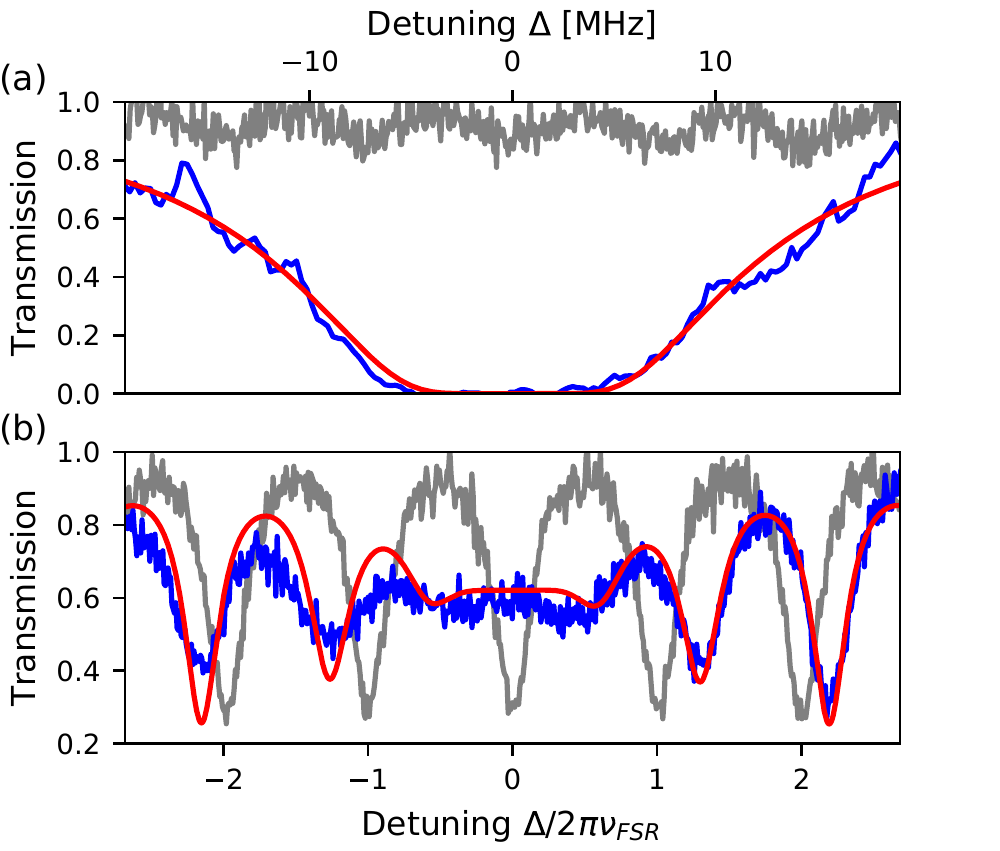}}
\caption{(a) Measured single-pass transmission spectrum of a MOT-cloud acquired through the TOF with the fiber-loop being open (blue line). The atomic cloud is optically dense in a frequency range of about $\pm 10$~MHz around the atomic resonance with a single-pass on-resonance optical depth of $\mathrm{OD} =12.7$, deduced from a fit with a saturated Lorentzian (red line). The transmission without atoms (grey line) is close to unity. The residual modulation is reminiscent of the cavity resonances because the resonator is not perfectly overcoupled. (b) Measured (blue line) and calculated spectrum (red line) of the coupled atom--resonator system. In comparison to the empty resonator spectrum (grey line), the central resonance is split and higher-order resonances are shifted outwards -- a clear signature of multimode strong coupling. The theory prediction relies on $\kappa_{0}/2\pi = 0.39$~MHz and $\kappa_{ext}/2\pi = 0.21$~MHz, which are deduced from the empty resonator spectrum, and the collective coupling strength, $g_N/2\pi = 12.2$~MHz $= 1.3 \times \nu_{\textrm FSR}$, which is deduced from the OD measured in (a). The atom--resonator detuning, $\Delta_\text{at}=\omega_\text{at}-\omega_\text{res}$ , is left as a free fit parameter which is found to be $\Delta_\text{at}/2\pi=150$~kHz.
}
\label{fig.spectra}
\end{figure}

We measure transmission spectra of the coupled atom--resonator system with open and closed fiber-loop. The probing phase consists of two measurement intervals of 1 ms duration, a first interval with atoms and a reference interval without atoms. The probe power is chosen such that the mean intra-cavity photon number is $<1$. Within each interval, the probe field's frequency is swept over 40~MHz  (about five FSRs) with an acousto-optical modulator. The transmission spectra are acquired using a single-photon counting module. Figure \ref{fig.spectra}(a) and (b) present experimental transmission spectra with the fiber-loop being open and closed, respectively. With the open fiber-loop (a), the transmission is inhibited by the atomic cloud over a frequency range of more than 10 MHz. The fit shown in red yields an on-resonance optical depth of  $\mathrm{OD} = 12.7$. For the closed fiber-loop (b), without atoms, the transmission spectrum shows the equidistant resonances of the empty cavity, see grey line. For the coupled atom--resonator system, the measured spectrum (blue line) qualitatively differs from both the empty resonator and the single-pass transmission spectrum of the atoms. In particular, the central resonance is split and higher-order resonances are shifted outwards. The theoretical prediction (red line, see \cite{SuppMat}) only contains a residual atom--resonator detuning, $\Delta_\text{at}=\omega_\text{at}-\omega_\text{res}$, as a free fit parameter. The latter is found to be $\Delta_\text{at}/2\pi=150$~kHz. The underlying collective coupling strength, $g_N/2\pi = 8.7$~MHz, was deduced from the OD measured in Fig.~\ref{fig.spectra}(a). The very good agreement between theory and measurement, for $g_N$ well in excess of the FSR, clearly reveals that our system operates in the multimode strong coupling regime. The slight deviation between theory and measurement can possibly be attributed to the detuning-dependent dipole forces exerted by the probe light in the resonator which are known to modify the atomic density near the nanofiber surface~\cite{Sague2007}. Given the fact that the atom--resonator detuning is much smaller than collective coupling strength, the single-atom decay rate, and the FSR, we set $\Delta_\text{at}=0$ in what follows.

\begin{figure}[h!]
\centerline{\includegraphics[width=1\columnwidth]{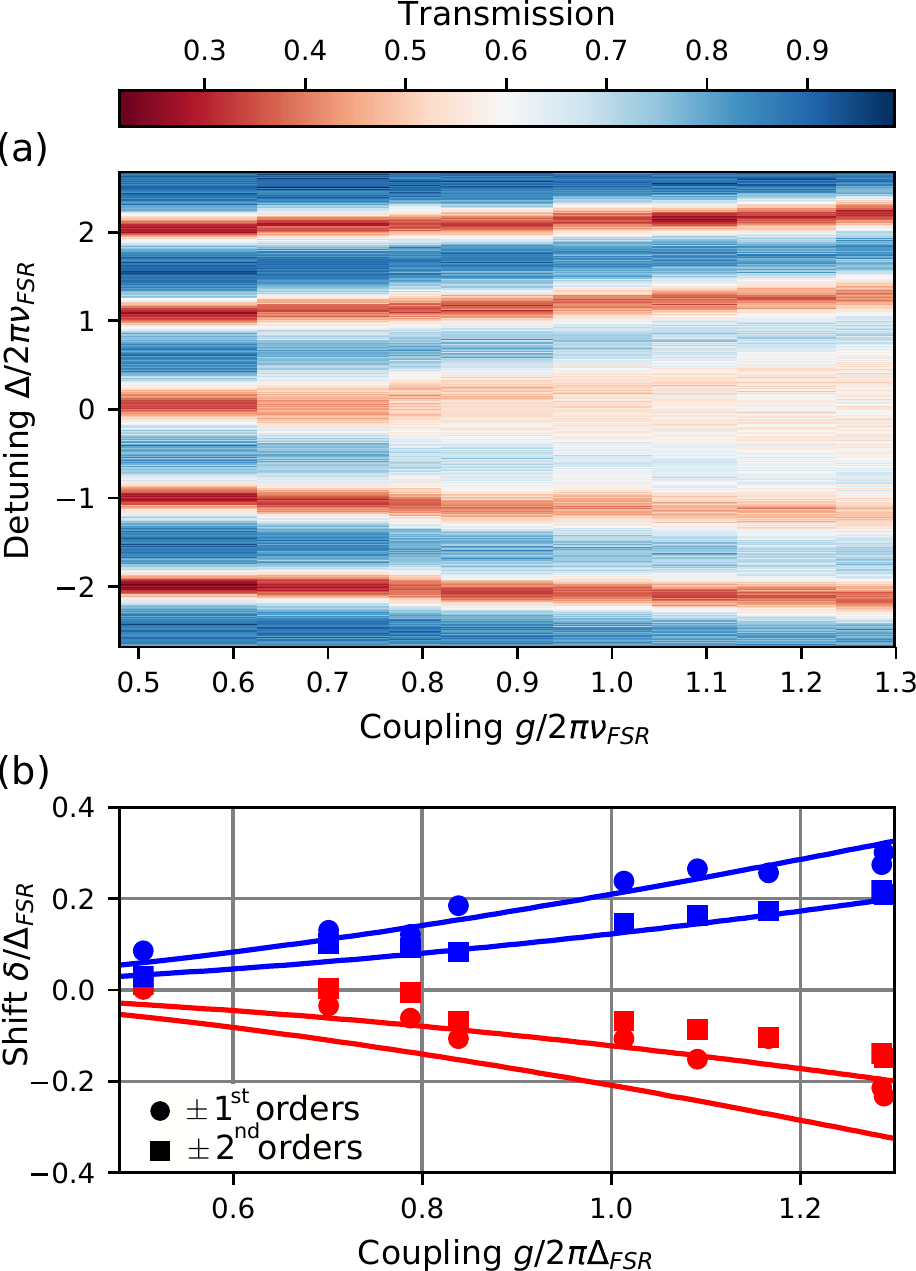}}
\caption{(a) Transmission of the atom--resonator system with increasing collective coupling strength, $g_N$.  The abscissa values of  $g_N$ are deduced from fitting the individual spectra. A splitting of the central resonance and an outward shift of the adjacent resonances that increase with $g_N$ are clearly visible, thus highlighting the multimode nature of the strong coupling. (b) Frequency shift of the transmission minima with respect to the empty cavity resonances. The red and blue squares and circles mark the positions of the transmission dips extracted from the experimental spectra. The solid lines give the theoretical prediction for the positions of transmission minima. The circles represent the $\pm 1^\text{st}$-order modes, whose bare-resonator frequency differs by $\pm \nu_\text{FSR}$ from the central resonance, and the squares the $\pm 2^\text{nd}$-order modes. As expected, both pairs of modes show a substantial shift on the scale of $\nu_\text{FSR}$, where the magnitude of the shifts of the $\pm 1^\text{st}$-order modes is larger than for the $\pm 2^\text{nd}$-order modes.}
\label{fig.TOFspectra}
\end{figure}

Figure \ref{fig.TOFspectra}(a) shows a series of transmission spectra in dependence of $g_N$. The latter is set by varying $N$ via the waiting time between switching off the MOT and probing the atom--resonator system. The abscissa values of  $g_N$ are deduced from fitting our model to the individual experimental spectra,  see \cite{SuppMat}. For $g_N/(2\pi\nu_\text{FSR})\gtrsim 1$, a splitting of the central resonance becomes visible and increases with $g_N$. Moreover, the outer resonances are progressively shifted outwards, and become shallower. The maximum coupling found here is $g_N/2\pi=(9.2\pm 0.07)$~MHz, in good agreement with the independently measured value from Fig.~\ref{fig.spectra}. Figure \ref{fig.TOFspectra}(b) shows the frequency shift, $\delta$, of the $\pm 1^\text{st}$-order and the $\pm 2^\text{nd}$-order resonator modes from to their original positions in the empty resonator as a function of $g_N$. Here, the magnitude of the shift of each mode increases with coupling strength: the $\pm 1^\text{st}$-order modes (shown in circles) shift more strongly than the  $\pm 2^\text{nd}$-order modes (in squares) which are further detuned from the atomic transition. The solid lines give the theoretical prediction for the positions of the modes. The agreement with the theoretical prediction is very good for the $+ 1^\text{st}$- and $+ 2^\text{nd}$-order modes. For the $- 1^\text{st}$- and $- 2^\text{nd}$-order modes, the observed shifts are smaller than expected from theory. We attribute this discrepancy to the detuning-dependent dipole forces exerted by the probe light, see above.

\begin{figure}[h!]
\centerline{\includegraphics[width=1\columnwidth]{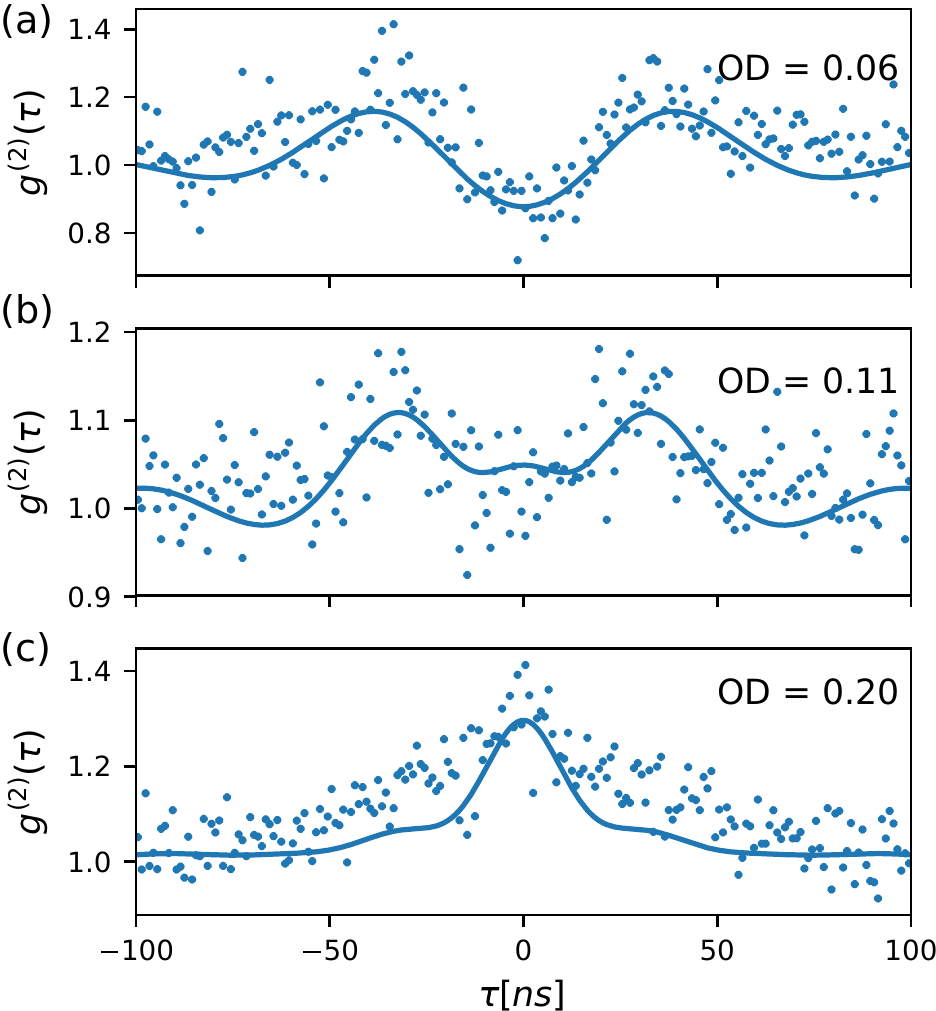}}
\caption{Second-order correlation measurements taken for three different ODs. We observe a transition from antibunching to bunching for increasing OD. Fitting the correlation data yields effective numbers of coupled atoms of 2.7, 4.3, and 9.5, respectively. From this data, we deduce the average optical depth per atom for our setup.}
\label{fig.at_num}
\end{figure}

In order to measure the number of coupled atoms, $N$, we excite the atomic cloud with the MOT laser beams and collect the resonance fluorescence photons that are emitted into the nanofiber-guided mode. For this measurement, the fiber-loop is open such that the light is directly guided towards the detector. The information on $N$ is then contained in the second-order (intensity-intensity) correlation function, $g^{(2)}(\tau)$. The measurement sequence alternates between a cooling phase (200~$\mu$s of magneto-optical trapping) and a 20-$\mu$s long fluorescence phase during which the MOT beams are switched into resonance with the atoms. In this setting, the form of $g^{(2)}(\tau)$ depends on the number of emitters contributing to the signal~\cite{Le_kien2008}. In particular, when many atoms radiate, $g^{(2)}(\tau)$ exhibits bunching at $\tau=0$, a signature originating from the stimulated emission processes involved. For small atom numbers, on the contrary, anti-bunching is expected at $\tau = 0$. Indeed, once a given atom emitted a photon, the emission of a second photon is prohibited on the timescale of the lifetime of the atomic state. This transition from anti-bunching to bunching happens at small atom numbers, which is the regime we placed ourselves in for this calibration experiment~\cite{Nayak2009, Grover2015}. Performing this measurement for  $\mathrm{OD}= 0.06 \pm 0.01$, $0.11 \pm 0.02$, and $0.20 \pm 0.01$, determined from independently measured transmission spectra, we obtain the correlation data shown in Figure \ref{fig.at_num}. 

We fit the model from~\cite{Le_kien2008} to the correlation data, see Supplemental Material, and find effective numbers of coupled atoms of $N_\text{eff}=2.7\pm0.3$, $4.3\pm0.3$ and $9.5\pm0.5$. The theory agrees well with the data but starts meeting its limits for $\mathrm{OD}= 0.20$. From $N_\text{eff}$ and the respective ODs, we deduce an average optical depth per atom of $\mathrm{OD} _{1} = 0.022 \pm 0.005$. This corresponds to an intrinsic single-atom cooperativity of  $C_1 = 0.26$, \cite{SuppMat}, a value which remains substantial despite the very long resonator. Using $\mathrm{OD} _{1}$, we can now convert any measured total OD into the effective number of coupled atoms. In particular, the threshold to MMSC is reached with as little as $N_\text{eff}=294 \pm 34$ atoms and the largest collective coupling of $g_N/2\pi=1.3\times\nu_\text{FSR}$ presented in Fig.~\ref{fig.TOFspectra} requires only $N_\text{eff}=645 \pm 109 $ atoms.

The experimental realization of multimode strong coupling for only 200 atoms sets the stage for devising novel applications and for the scientific exploration of cavity QED in an entirely new regime. For example, the expected non-Markovian dynamics of our experimental system is challenging theoretically and uncharted experimentally. In this context, pulsed revivals of the atomic inversion on the time scale of the round-trip time of the resonator, rather than sinusoidal vacuum Rabi oscillations, have been predicted \citep{Krimer2014}. Moreover, the interaction of different cavity modes that occurs under the condition of MMSC, in conjunction with the substantial single-atom cooperativity, should enable resonator-enhanced wave-mixing of fields containing only a few photons each. Finally, our system may also pave the way towards the implementation of quantum annealing algorithms using atoms~\cite{Torggler2017} or photons~\cite{Mcmahon2016} as carriers of information.\\

\textbf{Acknowledgements}\\

We thank Helmut Ritsch, Stefan Rotter, and Matthias Sonnleitner for helpful discussions. Financial support from the European Research Council (CoG NanoQuaNt) and the Austrian Science Fund (FWF, CoQuS project No. W 1210-N16 and NanoFiRe grant project No. P 31115) is gratefully acknowledged.\\

\bibliography{MMSC,citesm}

\clearpage

\section*{Supplemental Material}

\textbf{Transmission Model}\\

In the following section, we outline the derivation of the transmission spectrum of a multimode cavity. The considered setup consists of a coupling fiber and fiber-loop of length $L$ which constitutes the resonator. The coupling between the resonator and incoming fiber is realized using a fiber beamsplitter with an adjustable splitting ratio. The beamsplitter connects the coupling fiber input field $c_x^{\dagger}$ ($x\in[-\infty,0]$) and output field $d_x^{\dagger}$ ($x\in[0,+\infty]$) to the intraresonator field in the fiber loop $a_l^{\dagger}$ ($l\in[0,L]$). Here, $c_x^{\dagger}$ and $d_x^{\dagger}$ are the creation operators for a photon at position $x$ incoming or outgoing of the resonator and $a_l^{\dagger}$ is the creation operator for a photon at position $l$ in the resonator. Note that we assume uni-directional coupling (i.e. no backscattering) of the atoms and thus only consider the forward propagating field modes. The positions $x=0$, $l=0$, and $l=L$ denote the different ports of the beamsplitter. Using these definitions and following the approach outlined in \cite{ShenFan} the full problem can be described by the Hamiltonian:
\begin{equation}
\begin{split}
\frac{H}{\hbar} = & \int_{-\infty}^0dx\int_0^{+\infty}dx'\int_0 ^Ldl\left[c_x^{\dagger}\left(-iv_g \frac{\partial}{\partial x}\right)c_x \right.\\
& + d_{x'} ^{\dagger} \left( -iv_g \frac{\partial}{\partial x} \right) d_{x'} + a_{l} ^{\dagger} \left( -i v_g \frac{\partial}{\partial x} \right) a_{l} + U_{bs} \\
& + \sum_{n = 0} ^N \left(\omega_a(\sigma_n ^+ \sigma_n ^- - i\gamma) \right. \\
& + \left. \left. \delta(l - l_n) V(\sigma_n ^+ a_{l} + \sigma_n ^- a_{l}^\dagger) \right) \right],
\end{split}
\end{equation}
where $\sigma^+_{n}$ ($\sigma^-_{n}$) are the raising (lowering) operators for atom $n$ at position $l_n$, $N$ the total number of atoms, $v_g$ the group velocity of light in the resonator, and $\omega_a$ the atomic resonance frequency. Furthermore, $V = \sqrt{2 v_g \beta \gamma}$, where $\beta=\gamma_{nf}/\gamma$ is the ratio of the spontaneous emission rate of the atom into the nanofiber mode $\gamma_{nf}$ and the total emission rate $\gamma$. The (non-Hermitian) operator $U_{bs}$ describes the coupling between the resonator and coupling fiber and is given by:

\begin{equation}
U_{bs}/v_g = t_{rt}(i t_1  a_La^\dagger_0+ t_2 a_Ld^\dagger_0) +i t_1 c_0d_0^\dagger + t_2 c_0a^\dagger_0,
\end{equation}

where $t_1$ is the transmission coefficient of the beamsplitter for the transformations $c^\dagger_0\rightarrow d^\dagger_0$ and $a^\dagger_L\rightarrow a^\dagger_0$ and $t_2$ is the transmission coefficient for $c^\dagger_0 \rightarrow a^\dagger_0$ and $a^\dagger_L\rightarrow d^\dagger_0$. The factor $t_{rt}$ accounts for resonator roundtrip losses. The transmission coefficients are related to the standard resonator parameters via $t_{rt} = \sqrt{1-\kappa_0/(2\pi\nu_{FSR})}$ for the single round trip transmission of the resonator, and $t_1 = \sqrt{1-\kappa_{ext}/(2\pi\nu_{FSR})}$.

To obtain the steady state solutions we have to solve the eigenvalue problem $H|\Psi\rangle=\epsilon|\Psi\rangle$. For a single photon the general quantum state is given by

\begin{equation}
\begin{split}
|\Psi\rangle=&\left[\int_{-\infty}^{0}dx\phi_c(x)c^\dagger_x+\int_{0}^{\infty}dx'\phi_d(x')d^\dagger_x+\right.\\
&\left.\int_{0}^{L}dl \phi_a(l) a^\dagger_l\right]|0\rangle.
\end{split}
\end{equation}
Without loss of generality, we assume that the atomic ensemble is located at position $l_0$ inside the resonator. We can then make the Ansatz that the steady state photon wavefunction is described by $\phi_a(l) = \phi_{a_1} e^{ikl}$ for $l\in[0,l_0]$, $\phi_a(l) = \phi_{a_2} e^{ikl}$ for $l\in[l_0, L]$, $\phi_c(x) = \phi_c e^{ikx}$ for $x\in[-\infty,0]$, and $\phi_d(x) = \phi_d e^{ikx}$ for $x\in[0,+\infty]$. Here, $k = \omega/v_g$ with $\omega$ the frequency of the light.

From this we obtain the following set of equations for the amplitudes $\phi_i$ in the steady-state case:

\begin{equation}
\begin{split}
0 = & - \phi_{a_1} + t_2 \phi_c + i t_{rt} t_1 \phi_{a_2}\\
0 = & - \phi_d + it_1 \phi_c + i t_{rt} t_2 \phi_{a_2}\\
0 = &\,\,t_{at}^N \phi_{a_1} - \phi_{a_2},
\end{split}
\end{equation}
where $t_{at}$ describes the transmission through a single atom given by
\begin{equation}
t_{at} =  \frac{\gamma-\beta\gamma-i\Delta}{\gamma + \beta\gamma-i\Delta},
\end{equation}
with $\Delta = \omega_a - \omega$ the atom-light detuning. Solving this set of equations, we obtain for the steady-state transmission through the fiber:

\begin{equation}
T=\left| \frac{\phi_d}{\phi_c}\right|^2= \left| \frac{ e^{ikL} t_{rt} t_{at}^N - t_1}{ e^{ikL} t_{rt} t_{at}^N t_1 - 1} \right|^2,
\label{eq.transm_coeff}
\end{equation}

where $kL = \omega/\nu_{FSR}$.\\

\textbf{Relation between OD and $g_N$}\\

To deduce the collective coupling strength $g_N$, we can make use of a related experimentally accessible quantity, i.e. the optical depth OD of the atomic ensemble coupled to the nanofiber. For uni-directional coupling, the optical depth of a single atom (OD$_1$) is related to the fraction of spontaneous emission into the nanofiber-mode $\beta$ by $\text{OD}_1 = 4\beta$, provided $\beta \ll 1$. Moreover, the cooperativity of the ensemble of atoms is directly related to the single-pass optical depth of the atomic cloud and the resonator finesse $\mathcal{F}$, so that:

\begin{equation}
C_{N} = \frac{1}{\pi} \mathcal{F} \cdot  \text{OD},
\end{equation}
Alternatively, the cooperativity is also defined by
\begin{equation}
C_{N} = \frac{g_N^2}{2 \kappa \gamma}.
\end{equation}
Combining these two expressions, the collective coupling $g_N$ can be expressed as 
\begin{equation}
g_N^2 = 2 \, \nu_{FSR} \, \gamma \, \text{OD}.
\end{equation}

The optical depth of the cloud therefore gives a measurement of the collective coupling. This quantity is easily accessed in our experiment \cite{Vetsch2010}.\\

\textbf{Locking of the cavity}\\

To lock the resonator to the atomic transition, we use polarization spectroscopy as detailed in \cite{Libson2015}. We feedback onto a piezoelectric element, stretching the fiber resonator in response to external perturbations. The experimental sequence alternates between probing and locking. An acousto-optical modulator (AOM) is used to switch between the detection setup and the probing setup. The main advantage of the AOM compared to a 50/50 beamsplitter is that very little probe beam power is lost in the detection chain. \\

The lock's bandwidth is about 100 Hz. Our resonator exhibits some residual fluctuations at $\sim$140 Hz, leading to a broadening of the resonator lines. These fluctuations are most probably related to a mechanical excitation of the fiber mount. In order to compensate for this, we apply a recentering algorithm to overlap data from different times. This algorithm relies on the reference spectrum of the unloaded cavity which we acquire 3 ms after recording each shot with atoms. All the empty resonator spectra are overlapped by minimising the least square differences and the resulting correction is applied to the loaded resonator spectra.\\

\textbf{Correlation measurements of resonance fluorescence: OD per atom calibration}\\

In the following, we detail how we derive the number of atoms in the ensemble interacting with the resonator mode from the second-order correlation function. We set the resonator to the fully overcoupled condition and excite the atoms using the beams of the magneto-optical trap, set to be resonant to the Cs cycling transition. The fraction of fluorescence photons scattered into the nanofiber is collected on two detectors placed after a 50/50 beamsplitter. These measurements are conducted at intensities exceeding the saturation intensity of the atomic transition to collect sufficient photon numbers, and were performed for different optical depths. The temporal shape of the measured correlations then contains information on the atom number \cite{Le_kien2008, Nayak2009, Grover2015}. The non-normalized second order correlation function $G^{(2)}$ for the light scattered by the atomic ensemble is given by~\cite{Le_kien2008}
\begin{equation}
G^{(2)} \propto N \Gamma^{(2)}(\tau) + N (N-1) \left[ \mu_0 + \mu | \Gamma^{(1)}(\tau) |^2 \right],
\label{eq.G2}
\end{equation}
where $\tau$ is the delay between two photon-detection events, $N$ is the number of atoms in the ensemble and $\Gamma^{(1)}$ ($\Gamma^{(2)}$) the single-atom first- (second-) order correlation function. These two functions are analytically known \cite{Le_kien2008} and depend on the Rabi frequency of the excitation light. The coefficients $\mu_0$ and $\mu$ are geometric parameters taking into account the profile function of the nanofiber-guided modes. For small atom numbers, the first-order term dominates and gives rise to anti-bunching: if a single atom emitted a photon, the emission of a second photon is inhibited on a time scale of the lifetime of the probed transition. On the other hand, for large atom numbers the term proportional to $\Gamma^{(2)}$ dominates, and stimulated emission leads to bunching.\\

To obtain the atom number, we fit Eq.(\ref{eq.G2}) to our measured data. We use a constrained fit, where the values for $\mu$, $\mu_0$ are restricted to intervals taken from \cite{Nayak2009}. These quantities are not known exactly in the case where atoms are randomly distributed around the nanofiber with random dipole orientations. The error due to this uncertainty dominates the error on the determined atom number. The Rabi frequency is left as a free fit parameter. We found similar values for the three sets of data, which were taken in the same experimental conditions. The background is subtracted and the measured correlations are normalized by taking the asymptotic values reached at 10~$\mu$s, much longer than the typical time scales at play here. This model agrees well with the measured data for small atom numbers.

\end{document}